\documentclass{iau} 

\usepackage{graphicx}
\usepackage{xcolor}
\usepackage[hyphens]{url}
\usepackage{hyperref}
\hypersetup{
     colorlinks  = true,
     linkcolor   = blue, 
     anchorcolor = BrickRed,
     citecolor   = blue,
     filecolor   = blue,
     menucolor   = blue,
     runcolor    = cyan,
     urlcolor    = blue
}
\usepackage[authoryear]{natbib}

\pubyear{2019}
\volume{355} 
\jname{The Realm of the Low-Surface-Brightness Universe}
\editors{D. Valls-Gabaud, I. Trujillo \& S. Okamoto, eds.}

\def\etal{\textit{et al.}}

\title[The GSMF from CCSNe with improved photo-$z$ techniques] 
{The galaxy stellar mass function from \\ CCSNe with improved photo-$z$ techniques}

\author[T. M. Sedgwick, I. K. Baldry, P. A. James \& L. S. Kelvin]  
{T. M. Sedgwick$^{1}$, I. K. Baldry$^{1}$, P. A. James$^{1}$
\and L. S. Kelvin$^{1,2}$}
\affiliation{$^{1}$Astrophysics Research Institute, Liverpool John Moores University, IC2, Liverpool Science Park, 146 Brownlow Hill, L3 5RF
\\[\affilskip]
$^{2}$Department of Astrophysical Sciences, Princeton University, 4 Ivy Lane, Princeton, NJ 08544, USA}

\begin{document}

\maketitle

\begin{abstract}
In Sedgwick et al.\ (2019) we introduced and utilised a method to combat surface brightness and mass biases in galaxy sample selection, using core-collapse supernovae (CCSNe) as pointers towards their host galaxies, in order to: (i) search for low-surface brightness galaxies (LSBGs); (ii) assess the contributions of galaxies at a given mass to the star-formation-rate density (SFRD); and (iii) infer from this, using estimates of specific star-formation (SF) rate, the form of the SF-galaxy stellar mass function (GSMF). A CCSN-selection of SF-galaxies allows a probe of the form of the SFRD and GSMF deep into the dwarf galaxy mass regime. In the present work, we give improved constraints on our estimates of the SFRD and star-forming GSMF, in light of improved photometric redshift estimates required for estimates of galaxy stellar mass. The results are consistent with a power-law increase to SF-galaxy number density down to our low stellar mass limit of $\sim10^{6.2}$ M$_{\odot}$. No deviation from the high-mass version of the surface brightness -- mass relation is found in the dwarf mass regime. These findings imply no truncation to galaxy formation processes at least down to $\sim10^{6.2}$ M$_{\odot}$.
\end{abstract}

\firstsection 

\section{Introduction}\label{sec:intro}

The galaxy stellar mass function (GSMF) is a crucial probe of galaxy formation and evolution processes. Constraining present-epoch number densities of dwarf galaxies over a unit cosmological volume is of particular interest to present-day extra-galactic astronomy, as predictions of these densities from both semi-analytical and hydro-dynamical simulations \citep{MOO99,SCH15,GEN14} are particularly sensitive to the input cosmology and feedback prescriptions.

We discuss in \citet[][henceforth, S19a]{S19a} that the `sub-structure problem', or the the persistent under-abundance of observed dwarf galaxies relative to predicted numbers from simulations invoking $\Lambda$-CDM, may arise due to surface brightness biases in observational galaxy samples. As low-mass galaxies are typically fainter in surface brightness, they are more likely to be missed by source detection pipelines, and the observed dwarf population is expected to be incomplete.

S19a presented a method designed to mitigate against these surface brightness biases using a sample of core-collapse supernovae (CCSNe) from the SDSS-II Supernova Survey \citep{SAK18} as star-forming (SF) galaxy selection tools. CCSNe peak at luminosities of up to {$10^{8} - 10^{9}$ $L_{\odot}$}. They can be of higher luminosity than their host galaxies, and so can be used as pointers to their hosts. Whilst most of the 2456 SN-hosts were previously known to galaxy surveys, 140 dwarf galaxies were found for the first time at the locations of the SNe, in IAC Stripe 82 legacy images \citep[henceforth, `Stripe 82';][]{FT16}. 

From estimates of redshifts, host galaxy stellar masses ($\mathcal{M}$), CCSN selection, and associated uncertainties, CCSN-rates as a function of mass were derived.
This yields one of the most direct measurements of cosmic SF-rate density (SFRD).
From SFRD as a function of mass, an assumption for specific-SFR (sSFR) leads to estimates of the form of the GSMF of SF galaxies. A CCSN-selection of galaxies enables a probe of the GSMF deep into the dwarf regime of mass.

These results rely on an accurate calculation of galaxy stellar masses from photometry and redshifts. A large fraction of the low surface brightness galaxies (LSBGs) in the S19a sample possess only photometric measurements. We therefore rely on photometric redshift (photo-$z$) techniques when estimating galaxy stellar masses for the majority of the low-surface brightness galaxy sample.

S19a introduced and utilised zMedIC as a simple but useful photo-$z$ estimator relying on SDSS $ugriz$ photometry and a training set of Stripe 82 galaxies with spectroscopy. In the present work, we use `scaled flux matching' as an alternative photo-$z$ method in order to test for the effects on constraints of CCSN-rates and galaxy number densities in the dwarf regime. Significant improvements to galaxy stellar mass estimates arising from photo-$z$ method developments could help yield important insights into galaxy evolution.

\section{Updated Methodology}\label{sec:method}
This section outlines improvements to the methodology presented in S19a for the calculation of galaxy {photo-$z$'s} and their associated uncertainties. 

\subsection{Photometric redshifts using `scaled flux matching'}

CCSN hosts were identified in Stripe 82 data \citep{FT16} as discussed in Section \ref{sec:intro} and in greater detail in S19a. For SN hosts lacking spectroscopy from either the host galaxy or from the SN itself, we require a {photo-$z$}. We first formed a reference sample using galaxies from the IAC Stripe 82 catalogue that have a match to a spectroscopic redshift (spec-$z$) within $2.5''$.

The {photo-$z$'s} for the SN-host and other Stripe 82 galaxies were obtained 
by scaled flux matching (SFM). 
The is an empirical method that specifically works in linear flux space 
(unlike e.g.\ \citealt{BEC16} that uses colours and magnitudes) 
and requires a matching set of galaxies with reliable {spec-$z$'s}. 
The matching set were chosen to have the SDSS flag \textsc{zwarning} $=0$, $0.002 < z < 1.226$, 
reasonable S/N and reasonable measured colours for the redshift range. 
(S/N$ > 10$ in $r$ and $i$ bands, and S/N$ > 5$ in $g$ and $z$ bands,
colours in the range $-0.1 < u-g < 3.5$, $-0.1 < g-r < 2.2$, $-0.1 < r-i < 1.5$, $-0.1 < i-z < 1.0$.)
The matching set consisted of 103\,376 galaxies out of a total sample 
of 117\,690 (Stripe 82 catalogue with SDSS redshifts and SN hosts, removing duplicates).

For each source ($i$), the fluxes were fitted to matching-set galaxy ($j$) fluxes with $\chi^2$ defined as: 
\begin{equation}
  \chi^2_{i,j} = \sum_k \frac{(f_{i,k} - n_{i,j} f_{j,k})^2}{\sigma_{i,k}^2}
\end{equation}
where $n_{i,j}$ is the best-fit normalisation 
and the summation is over all the photometric bands ($k$). 
The photometric errors $\sigma_{i,k}$ were taken to be the linear 
flux errors from \textsc{SExtractor} added in quadrature to fractional errors: 
$0.02 f_{i,k}$ for $g,r,i,z$ and $0.05 f_{i,k}$ for the $u$-band.  
Note aperture fluxes with radius $2.4''$ were used. 

The reliability weight of the match is then given by: 
\begin{equation}
w_{i,j} = \exp( -\chi^2_{i,j} / 2) \, W_{j}
\end{equation}
where $W_{j}$ is the bin weight assigned to galaxy $j$ of the matching set. 
This was obtained by binning by 0.002 in $\zeta = \ln(1+z)$ (over the range 0.002--0.8) and 
with galaxies in a bin given a weight of $1/n_{\rm bin}$ with a maximum weight of 1/25. 
Note also that $w_{i,j}$ is set to zero where 
$i$ and $j$ refer to the same galaxy for calibration purposes. 

The weighted mean of $\zeta$, for the best-estimate {photo-$z$}, is given by:
\begin{equation}
  \zeta_{i,\mathrm{phot}} = 
  \frac{\sum_j w_{i,j} \zeta_{j,\mathrm{spec}}}
       {\sum_j w_{i,j}}
\end{equation}
where $\zeta$ is the appropriate quantity 
to use when dealing with redshift measurements and errors \citep{BAL18}. 
The initial estimate of the uncertainty (or the nominal error) is given by: 
\begin{equation}
  \zeta_{i,\mathrm{err}}^2 
  = \left( \frac{\sum_j w_{i,j} \zeta_{j,\mathrm{spec}}^2}
         {\sum_j w_{i,j}} \: - \:  \zeta_{i,\mathrm{phot}}^2 \right)
    \frac{N_{i,\mathrm{eff}}}{N_{i,\mathrm{eff}} - 1} 
\end{equation}
which is the weighted standard deviation multiplied by a correction 
factor to obtain the sample standard deviation, 
and $N_{i,\mathrm{eff}}$ is the effective number of measurements for reliability weights. 

\subsection{An improved assessment of photo-$z$ systematics}\label{sec:wPDF}

S19a presents in detail the methodology used to assess {photo-$z$} uncertainties for the CCSN-host galaxy sample, but in brief, we sub-divided the matching set of galaxies by {photo-$z$} into bins of width 0.025, before determining the {spec-$z$} distribution for each bin. This yields a set of redshift probability density functions (PDFs) which can be drawn from a Monte Carlo (MC) technique for a given {photo-$z$} input, in order to calculate galaxy stellar masses and volume-limit the sample ($z<0.2$) for each iteration. A benefit of this technique is that it can account for any systematic offsets between {photo-$z$} and {spec-$z$}, and PDFs need not be Gaussian.

A shortcoming of the technique is that it does not account for errors on a galaxy-by-galaxy basis. To mitigate against this problem we modify our method as follows:

\begin{enumerate}
    \item For each {photo-$z$} bin, we rank galaxies by their nominal error estimated from the SFM technique.
    \item We then determine the {spec-$z$} distributions using galaxies from the 0th to 20th percentile, and from the 80th to 100th percentile. We denote the distributions as $\mathcal{P}_{10}(z)$ and $\mathcal{P}_{90}(z)$, respectively, which are normalised to integrate to unity.
    \item If a galaxy ($i$) has a mean nominal SFM redshift error at the ($j$)$^{th}$ percentile of the {SFM error-ranked} spectroscopic sample, then the PDF, $P_{i}(z)$, used as the galaxy's input into the MC-technique is given by: 
\end{enumerate}
\begin{equation}\label{eq:PDFs}
    P_{i}(z) = \frac{100-j}{100}\mathcal{P}_{10}(z) + \frac{j}{100}\mathcal{P}_{90}(z)
\end{equation}
This improvement to the method accounts for changes in {photo-$z$} vs {spec-$z$} space with the SFM nominal error.

\section{Results \& Discussion}\label{sec:results}

\subsection{Improved constraints on CCSN-rate densities \\and the star-forming galaxy stellar mass function}\label{sec:densities}

Improved estimates of {photo-$z$'s} and their uncertainties are in particular crucial for the constraining of CCSN-rates, star-formation-rates, and star-forming galaxy number densities of low stellar mass galaxies. This is firstly because high surface brightness objects, and therefore typically higher stellar mass objects are more typically the focus of spectroscopic analyses, and secondly because only 27 of 140 low-surface-brightness SN-host galaxies discovered in S19a have associated spectroscopy (those which do use SN {spec-$z$}). 

\begin{figure}
\begin{center}
 \includegraphics[width=0.69\textwidth]{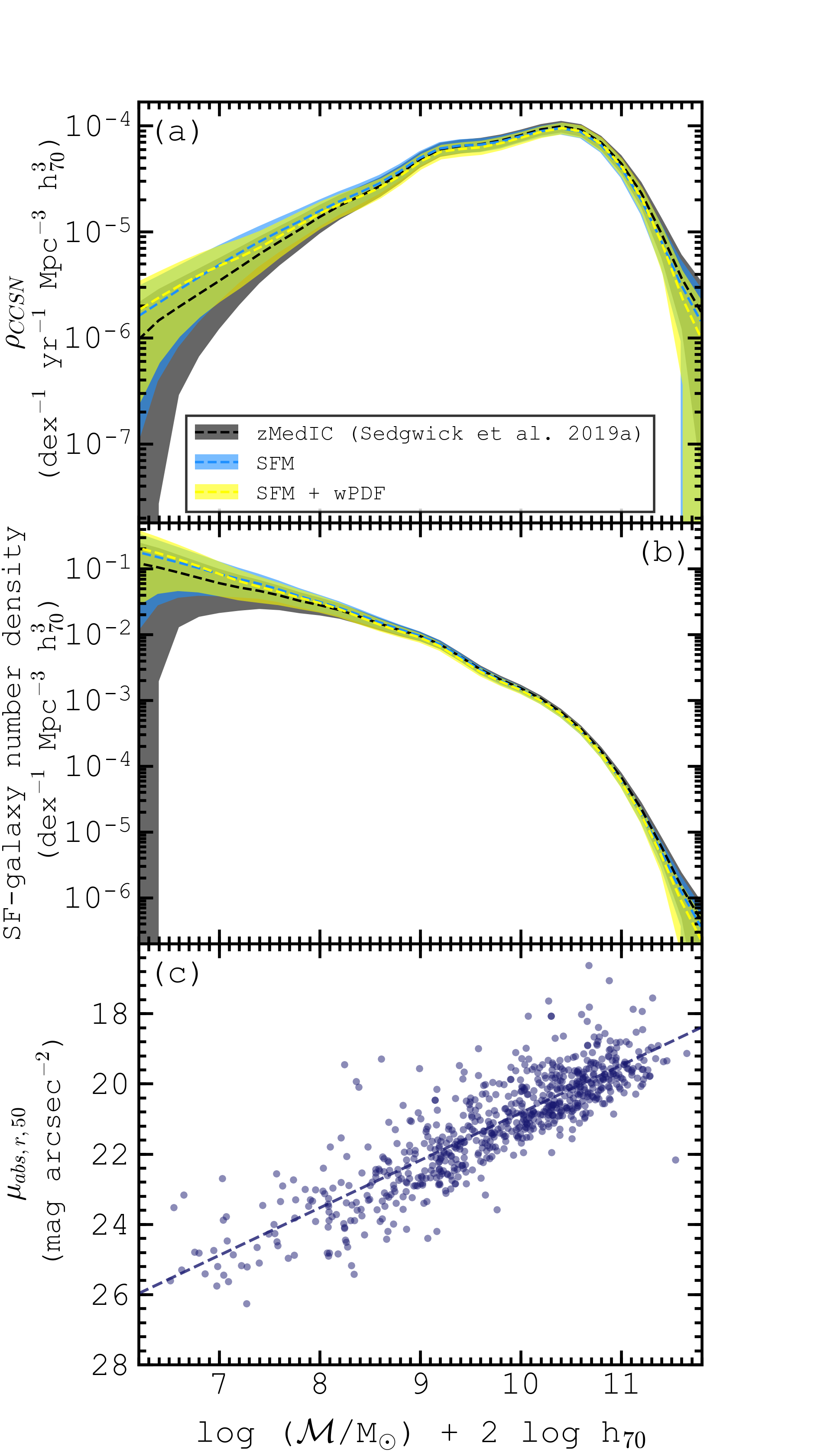}
 \caption{(a) Volumetric {$z < 0.2$} CCSN-rate densities as a function of galaxy stellar mass and (b) the {$z < 0.2$} star-forming GSMF; using 3 {photo-$z$} techniques: zMedIC, introduced in S19a (shown in black); SFM, shown in blue; and a modification of SFM using redshift PDFs dependant on the nominal redshift error estimate from SFM ({SFM + wPDF}; shown in yellow). (c) Effective absolute $r$-band galaxy surface brightness of CCSN host galaxies as a function of galaxy stellar mass using {SFM + wPDF}. The least-squares regression-line fit to the data is shown as the dashed line.}
 \label{fig:merged}
\end{center}
\end{figure}

Using different {photo-$z$} implementations, we calculate CCSN-rate densities as a function of galaxy stellar mass (corrected for the SN survey flux-limit; see S19a). These results are shown in the top-panel of Figure \ref{fig:merged}. All errors shown are the quadrature sum of MC + Poisson + cosmic variance errors. 
The result presented in S19a using the zMedIC {photo-$z$} method is shown in black. Using SFM to derive {photo-$z$'s}, shown in blue, the uncertainties on CCSN-rates in the dwarf galaxy regime of mass are significantly reduced. A slight increase to mean estimates of CCSN-rates in each bin over MC iterations is found below masses of {$\log (\mathcal{M}/\mathrm{M}_{\odot}) \lesssim 8.0$}. Accounting for nominal redshift error estimates from the SFM method, as discussed in Section \ref{sec:wPDF}, we note an additional reduction to CCSN-rate density uncertainties at low masses, shown in yellow (SFM + wPDF). This result gives further evidence for a power-law decrease to volumetric CCSN-rate densities with decreasing galaxy stellar mass. We are also able to probe 0.2 dex lower in galaxy stellar mass using the SFM technique compared to zMedIC.

Using the assumption of a constant sSFR, we use volume-integrated CCSN-rate estimates as a function of mass to estimate the SF-GSMF.
Binned CCSN statistics are divided by their host galaxy mass.

Values are then normalised to the \citet{BAL12} SF-galaxy number densities over the range {$8 < \log (\mathcal{M}/\mathrm{M}_{\odot}) < 9$} to yield estimates of the SF-GSMF from CCSNe, shown in the middle panel of Figure \ref{fig:merged}.

We highlight once more the significant improvement to constraints of densities in the dwarf galaxy mass regime when moving from zMedIC (shown in black) to SFM (shown in blue), with the fractional error on the number density a factor of $\sim 3$ smaller at {$\log (\mathcal{M}/\mathrm{M}_{\odot}) = 7.0$}. We suggest this improvement is due to the fact that zMedIC requires a single relationship between galaxy colours and redshift whereas SFM is flexible to spectral variety in the galaxy population and does not require a knowledge of the optimal parameterisation of colour versus redshift.

We show, in yellow, further constraints using the {SFM + wPDF} approach. Whilst these additional improvements are less
significant, the lower limit on the number density is a factor of $\sim 3$ higher at our low mass limit of {$\log (\mathcal{M}/\mathrm{M}_{\odot}) = 6.2$}. The trend is consistent with a power-law rise to SF-galaxy number densities continuing deep into the dwarf regime of mass.

This lack of deviation from a power law below {$\log (\mathcal{M}/\mathrm{M}_{\odot}) \sim 9.0$} implies there is no evidence for any truncation to galaxy formation processes down to at least {$\log (\mathcal{M}/\mathrm{M}_{\odot}) = 6.2$}. In Section \ref{sec:mu_mass}, we discuss further implications for galaxy formation processes in light of trends of galaxy surface brightness with stellar mass.

\subsection{The galaxy surface brightness -- stellar mass relation}\label{sec:mu_mass}

Star-forming galaxies exhibit a well-established trend of surface brightness with stellar mass \citep[see, e.g.][]{BAL12}, indicating that surface brightness scales with stellar mass. 
A break from this relation with decreasing mass could imply that star-formation is constrained by different processes below a certain mass.

The effective absolute $r$-band surface brightness of a CCSN host galaxy is computed using Equation \ref{eq:mu}, where $A_{r,50}$ denotes the area in arcsec$^{2}$ enclosing half the total galaxy flux, $m_{r,auto}$ is the Galactic extinction-corrected $r$-band \textsc{auto} apparent magnitude from \textsc{SExtractor}, $K_{rr}(z)$ is the k-correction using the prescription of \citet{CHI10}, and the final term is the cosmic surface-brightness-dimming correction.
\begin{equation}\label{eq:mu}
    \mu_{abs,r,50} = m_{r,auto} + 2.5 \log (A_{r,50}) + 2.5 \log (2) - K_{rr}(z) - 10 \log (1+z)
\end{equation}

The bottom panel of Figure \ref{fig:merged} shows the observed trend of effective surface brightness vs galaxy stellar mass. Shown are all galaxies hosting SNe classified as CCSNe or `Unknown' by \citet{SAK18} which have a mean redshift from the STY$+$wPDF method of {$z < 0.2$} over 1000 MC iterations. We use the mean galaxy stellar mass over these iterations. We observe that the surface brightness vs mass relation is linear for the full range of galaxy masses, with the regression line taking the form {$\mu_{abs,r,50} \simeq - 1.35 \log (\mathcal{M}/10^9 \mathrm{M}_{\odot}) + 22$}. This result implies no phase change in galaxy evolution or star-formation processes down to {$\log (\mathcal{M}/\mathrm{M}_{\odot}) = 6.2$}, in concordance with our GSMF results of Section \ref{sec:densities}.

\subsection{Summary \& Conclusions}

We have presented the `scaled flux matching' technique for the calculation of {photo-$z$'s}, along with developments to the {photo-$z$} methodology of \citet{S19a} designed to improve the assessment of photo-$z$ systematics. A coupling of these developments applied to the methodology of S19a produces improved galaxy stellar mass estimates of CCSN host galaxies, leading to tighter constraints on trends of volumetric CCSN-rates, star-formation-rates and star-forming galaxy number densities with stellar mass.

These results put constraints on CCSN-rates at low galaxy masses, which can be used to trace volumetric low-mass-galaxy star-formation-rates. Estimates of the form of the star-forming GSMF show evidence for a power law rise to star-forming galaxy number densities with decreasing galaxy stellar mass, with no evidence of a down turn in galaxy number densities towards lower masses. This result helps to relieve tension between the observed numbers of dwarf galaxies and those predicted from simulations invoking $\Lambda$-CDM, and implies a lack of truncation to galaxy formation processes down to at least {$\log (\mathcal{M}/\mathrm{M}_{\odot}) = 6.2$}.
Further evidence for this conclusion is found from trends of surface brightness with galaxy stellar mass, with no break from the higher mass relation observed down to the low-mass limits of this study.

Future high-cadence surveys with increased time-span, area and surface brightness depth 
will greatly increase the size of CCSN-selected galaxy samples and lead to the detection of vast numbers of dwarf galaxies. 
Using expectations for these parameters with LSST \citep{IVE13}, and comparing with the specifications and CCSN yield of the SDSS-II Supernova Survey \citep{SAK18}, LSST may be expected to identify $\sim 25$ times more CCSNe in a year for a 5000 deg$^{2}$ coverage in rolling cadence regions. The present work's results suggest that a similar study with LSST would yield
$\sim 500$ $z<0.2$ CCSNe yr$^{-1}$ in star-forming galaxies with masses of {$\log (\mathcal{M}/\mathrm{M}_{\odot}) < 7.0$}, statistics which would undoubtedly give important insights into dwarf galaxy evolution. The co-adding of images is crucial for low-surface brightness host identification. Based on expected exposure times per visit, we estimate that only 25 LSST images per SN region will be required to match the depth of IAC Stripe 82 legacy imaging. Further co-added imaging will bolster prospects for LSBG detection, helping to probe the form the GSMF to ever lower masses, giving crucial insights into galaxy formation and evolution processes deep into the dwarf regime.

\end{document}